\documentclass[journal,twoside,web]{ieeecolor}
\usepackage{hyperref}\hypersetup{hidelinks}
\usepackage{etoolbox}
\makeatletter\patchcmd{\thebibliography}{\section*{\color{black}}}{\section*{\color{black}References}}{}{}\makeatother
\makeatletter\patchcmd{\biographynophoto}{\vskip 4\baselineskip plus 1fil}{\vskip 0.25\baselineskip plus 1fil}{}{}\makeatother
\usepackage{subcaption}
\usepackage{array}
\usepackage{dblfloatfix}
\usepackage{placeins}
\usepackage{generic}
\usepackage{cite}
\usepackage{amsmath,amssymb,amsfonts}
\usepackage{graphicx}
\usepackage{textcomp}

\begin{document}
\title{Multi-Sensor Fusion for False-Alarm Reduction
by Timed Electromechanical Corroboration}

\author{Zixuan Fu,~\IEEEmembership{Student~Member,~IEEE,}
Anthony Raffaele,~\IEEEmembership{Student~Member,~IEEE,} and Unnikrishna Pillai,~\IEEEmembership{Fellow,~IEEE}
\thanks{Manuscript submitted 2026. This work received no external funding.}
\thanks{The authors are with the Department of Electrical and
Computer Engineering, Tandon School of Engineering, New York University, Brooklyn, NY
11201 USA (e-mail: zf2375@nyu.edu; ar6568@nyu.edu; up240@nyu.edu). Z. Fu is a Research
Intern; A. Raffaele is an undergraduate student.}}

\maketitle

\begin{abstract}
A physician checks the clinical electrocardiogram against other observations. A wearable monitor
has no physician. This paper shows that a mechanical sensor worn beside the electrocardiogram can
take over that check, and that the right added sensors lower false alarms further. A beat
is accepted only when a mechanical contraction follows at the delay learned for the wearer.
Beat-shaped artifacts injected into resting and exercise recordings passed the electrocardiogram
alone almost every time. A seismocardiogram or phonocardiogram as the second channel rejected
most while keeping nearly all true beats, strongly at rest and partly during exercise.
Shifting the mechanical signal in time removed the benefit. A detection criterion scores an added
sensor by the part of its response that the sensors in place cannot predict. With four sensors,
this score ranked the sets as their measured false-alarm rates did. With the two
sensors it selected, 6.2 percent of the artifacts passed while 95 percent of true beats were kept,
half the rate of the best single sensor. In integer arithmetic on a low-cost
microcontroller, the detector made the same decisions as the software. These rates come from injected artifacts, not
alarms in patients. Natural motion and clinical events remain untested.
\end{abstract}

\begin{IEEEkeywords}
Electromechanical corroboration, embedded processing, false-alarm reduction, multi-sensor fusion, wearable health monitoring.
\end{IEEEkeywords}

\section{Introduction}
\begin{figure}[!t]\centering
\includegraphics[width=0.88\columnwidth]{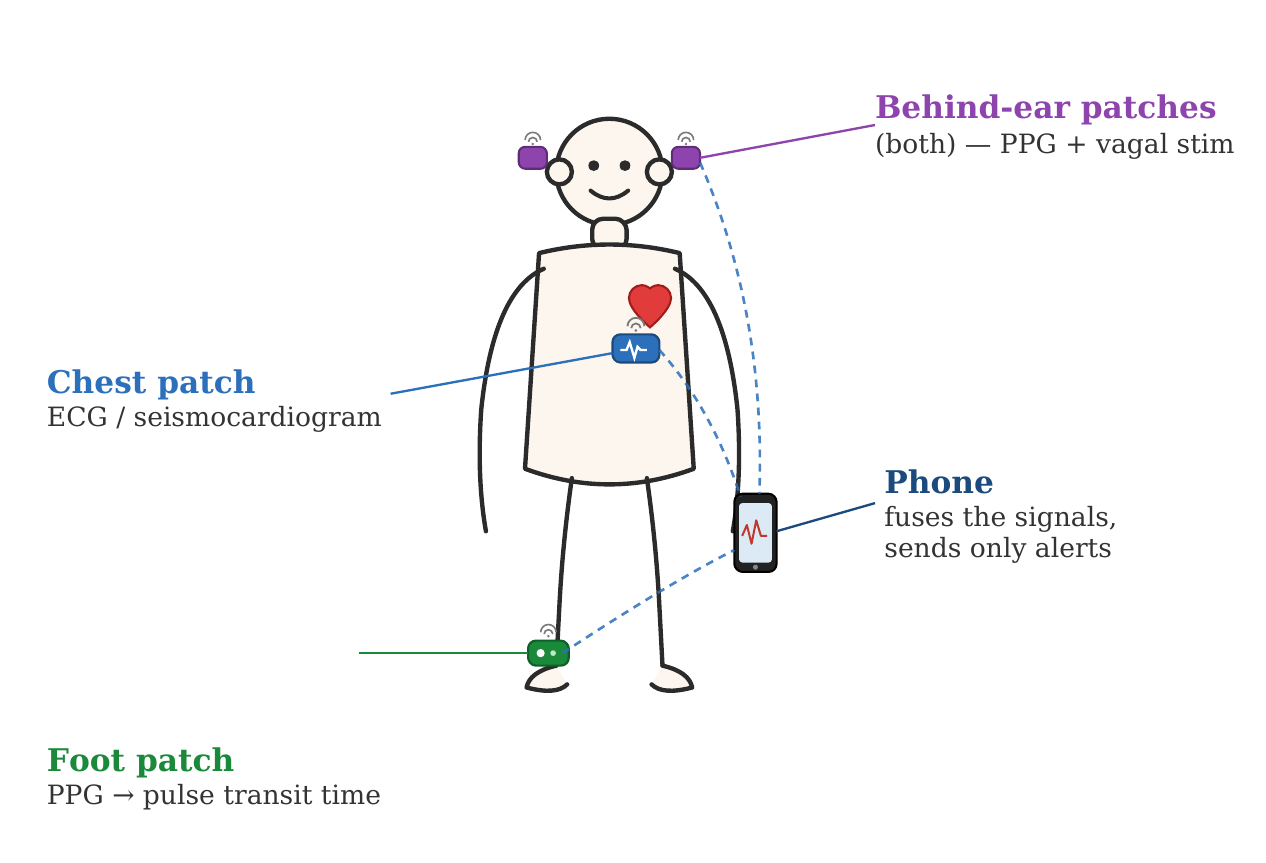}
\caption{Proposed phone-centered wearable architecture. The present experiments validate the
chest electrical and mechanical pair and phone-side fusion concept. The distal-pulse and
behind-the-ear sites are architectural extensions and are not experimentally validated here.}
\label{fig:overview}
\end{figure}
\IEEEPARstart{T}{he} electrocardiogram is trusted in the clinic because it is never read alone.
The patient lies still, the electrodes are placed and checked, and a physician judges each
complex against everything else in the room, the pulse under a finger, the sounds in a
stethoscope, and the look of the patient. The physician supplies the corroboration. An
artifact that looks like a beat on the trace is dismissed because nothing else agrees with it.
A wearable monitor has no physician. It carries one lead on a moving body and an algorithm that
accepts whatever has the energy and shape of a beat. Motion, poor contact, and modality-specific
artifacts can make that single channel appear abnormal when the physiology is not. Even at the
bedside, where the signal is far cleaner, the electrocardiogram alone fires false alarms often
enough that suppressing them is an open benchmark~\cite{clifford2015}. Frequent alerts lead users
to silence or ignore them, a problem known as alarm fatigue~\cite{shivers2013}, and the same risk
follows personal sensing into daily care~\cite{karter2026,natsensors2026}.

The aim of this paper is to replace the physician's corroboration with sensors and fusion. A
mechanical channel worn beside the electrocardiogram supplies what the physician's finger and
stethoscope supply, an independent response that a true beat produces and an electrical artifact
does not. Fusing that channel with the electrocardiogram lowers the false-alarm rate. Fusing
complementary channels lowers it further, and a detection criterion says which to add.
The physician's own fusion shows the two principles the method copies. The physician reads the
electrocardiogram, listens to the heart, feels the pulse, and trusts a finding only when these
observations agree. Each instrument fails in its own way, so agreement is stronger evidence than
any one reading. To locate a murmur, a physician also listens at the valve, moves the stethoscope
to a nearby spot to hear the background alone, and subtracts that background by ear. What the two
spots share cancels. Two sensors worn at once cancel a shared disturbance in the same way, without
the move.

Fusing channels to reject false events has a long history in bedside monitoring. The arterial
pressure waveform suppresses false critical electrocardiogram (ECG) alarms in intensive
care~\cite{aboukhalil2008}, and ECG signal-quality indices reduce false arrhythmia
alarms~\cite{behar2013}. The PhysioNet/Computing in Cardiology challenges of 2014 and 2015 made
multimodal beat detection and false-alarm reduction open benchmarks~\cite{silva2015,clifford2015}.
A common strategy takes each beat from whichever channel has the better signal
quality~\cite{johnson2015}. Wearable mechanical signals such as the seismocardiogram now bring
comparable information outside the clinic~\cite{inan2015}. The approach here does not choose the
better channel. It requires each electrical event to be confirmed by a mechanical event at the
delay learned for the subject, and it tests that timing directly with a time-shifted control.
Outside the clinic a cardiac emergency is often recognized only when a bystander notices it. A
wearable monitor could raise the alert earlier, but only if its alerts can be trusted. The
proposed architecture places the fusion on the phone (Fig.~\ref{fig:overview}). Low-cost sensors
stream locally, the phone forwards only the events that survive corroboration, and the high-rate
raw signals stay local. Phone-side deployment and network traffic are not measured here. Such a
monitor would raise alerts, not treat the wearer.

An electrical heartbeat candidate is more likely to be genuine when a mechanical contraction
follows at the subject's expected electromechanical delay. Three public multimodal resources test
that claim with a time-shifted control. One is at rest, one is during exercise, and one records
four mechanical sensors together. The third asks whether more sensors lower false admissions
further. A detection criterion answers that question before any operating curve is measured. In
a Gaussian model, the deflection of the joint confirmation features fixes the false-alarm probability
at a given detection probability. An added sensor is then worth the part of its response that the sensors
already in place cannot predict. The same detector runs on a microcontroller far more constrained
than a modern phone. In this controlled artifact challenge, an injected ECG-only transient has no
physiologically timed mechanical counterpart, so timing corroboration can reject it. The
experiments test three linked ideas. The first is physiological timing as a corroboration cue.
The second is the deflection increment as a way to select complementary sensors. The third is a
fixed-point implementation of the two-channel gate.

\section{Which Sensors to Combine}\label{sec:criterion}
Each candidate event yields one confirmation feature per mechanical sensor. For $M$ sensors the
features form a real vector $\underline{x}$ of length $M$, and a superscript asterisk denotes the
transpose. Under hypothesis $H_1$ the event is a true beat, and under $H_0$ it is an artifact. Let
$\underline{m}_0=E[\underline{x}\mid H_0]$ be the mean feature under an artifact and
$\underline{\mu}=E[\underline{x}\mid H_1]-\underline{m}_0$ the mean shift produced by a true beat.
The centered vector $\underline{z}=\underline{x}-\underline{m}_0$ then obeys
\begin{equation}
\underline{z}=\theta\,\underline{\mu}+\underline{w},\qquad \theta\in\{0,1\},
\end{equation}
with $\theta=1$ under $H_1$ and $\theta=0$ under $H_0$. The disturbance $\underline{w}$ is zero-mean
Gaussian with the $M\times M$ covariance $R=E[\underline{w}\,\underline{w}^{*}]$ under both
hypotheses. For two Gaussian densities with a common covariance the log-likelihood ratio is linear.
The most powerful test then compares $T(\underline{z})=\underline{\mu}^{*}R^{-1}\underline{z}$ with
a threshold $\gamma$. The statistic $T$ has variance $d^{2}$ under both hypotheses, mean zero under
$H_0$ and mean $d^{2}$ under $H_1$, where $d^{2}=\underline{\mu}^{*}R^{-1}\underline{\mu}$ is the
deflection. With $Q$ the Gaussian tail probability, the false-alarm probability is
$P_F=Q(\gamma/d)$ and the detection probability is $P_D=Q(\gamma/d-d)$, so that
\begin{equation}
P_F=Q\big(d+Q^{-1}(P_D)\big).
\end{equation}
In the experiments below, the retention of clean beats stands in for $P_D$ and the admission of
injected transients stands in for $P_F$. Neither is a clinical sensitivity or a clinical
false-alarm probability. Under this model, admission at a fixed retention falls strictly as $d$
grows. The set of sensors with the largest deflection therefore has the lowest predicted
admission at that retention. A timed mechanical feature has a
large mean shift because an ECG-only transient has no contraction at the learned delay. Reading the
mechanical channel at a wrong delay, as the time-shift control does, removes the mean shift and
with it the deflection.

The value of an added sensor follows from the same quantity. Partition the sensors into a set $A$
already in place and a candidate set $C$, with
\begin{equation}
\underline{\mu}=\begin{pmatrix}\underline{\mu}_A\\ \underline{\mu}_C\end{pmatrix},\qquad
R=\begin{pmatrix}R_A & R_{AC}\\ R_{CA} & R_C\end{pmatrix},\qquad R_{AC}=R_{CA}^{*}.
\end{equation}
The set $A$ alone has deflection $d^{2}(A)=\underline{\mu}_A^{*}R_A^{-1}\underline{\mu}_A$. Define
the unpredicted mean shift $\underline{\nu}=\underline{\mu}_C-R_{CA}R_A^{-1}\underline{\mu}_A$ and
the unexplained disturbance covariance $U=R_C-R_{CA}R_A^{-1}R_{AC}$, the Schur complement of $R_A$ in
$R$. The block inverse of $R$ gives
\begin{equation}
d^{2}(A\cup C)=d^{2}(A)+\underline{\nu}^{*}U^{-1}\underline{\nu}.
\end{equation}
The increment of the candidate is $\Delta d^{2}=\underline{\nu}^{*}U^{-1}\underline{\nu}\ge0$.
The vector $\underline{\nu}$ is the part of the candidate's mean shift that the sensors in place
cannot reproduce by linear regression on their disturbances, and $U$ is the part of its
disturbance that they cannot explain. The increment vanishes exactly when
$\underline{\mu}_C=R_{CA}R_A^{-1}\underline{\mu}_A$. In this Gaussian model such a sensor adds
nothing, however strong it is alone. For two sensors with unit disturbance variances, disturbance
correlation $\rho$ and mean shifts $\mu_1$ and $\mu_2$, the increment of the second sensor is
$(\mu_2-\rho\mu_1)^{2}/(1-\rho^{2})$. It is largest when the sensors share their disturbance but
not their signal, since the shared disturbance can then be subtracted.

\begin{figure}[!t]\centering
\includegraphics[width=0.72\columnwidth]{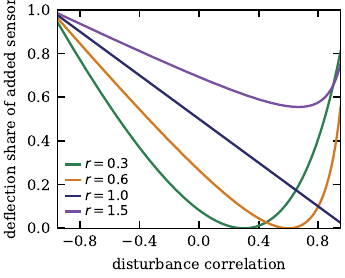}
\caption{Share of the two-sensor deflection contributed by the second sensor against the
disturbance correlation, for mean-shift ratios $r=\mu_2/\mu_1$ of 0.3, 0.6, 1.0 and 1.5.}
\label{fig:redundancy}
\end{figure}

Fig.~\ref{fig:redundancy} plots the share $\Delta d^{2}/d^{2}$ of the second sensor against $\rho$ for four ratios
$r=\mu_2/\mu_1$. For $r<1$ the share vanishes at $\rho=r$, where the mean shift of the second
sensor is exactly the regression of the first. For $r=1$ the share equals $(1-\rho)/2$. For $r>1$
it never vanishes, and its smallest value $1-1/r^{2}$ occurs at $\rho=1/r$. For every $r\ne1$ the
share rises toward one as $\rho\to1$, because a disturbance common to both sensors can be
subtracted and leaves the difference of their mean shifts exposed.
This increment is the classical measure of the additional discriminating information carried by a
group of variables~\cite{rao1973}. Applied to the confirmation features, it ranks the candidate
sensors before any operating curve is measured. The set stops growing when the increment of every
remaining sensor is near zero. Building the set one sensor at a time is not guaranteed to find the
best set of a given size. Two sensors that are weak alone can be strong together when their
disturbances are correlated and their mean shifts differ, as Fig.~\ref{fig:redundancy} shows near
$\rho=1$. The experiments below therefore evaluate every sensor set.

The same increment governs detection at a fixed false-alarm budget $P_{F0}$. Write $d_A=d(A)$ and
$q=Q^{-1}(P_{F0})$. The enlarged set detects with probability $Q\big(q-d(A\cup C)\big)$. For a
small increment, $d(A\cup C)\approx d_A+\Delta d^{2}/(2d_A)$, and with $\varphi$ the standard
Gaussian density,
\begin{equation}
\Delta P_D\approx \varphi(q-d_A)\,\Delta d^{2}/(2d_A).
\end{equation}
The result is exact for the stated model only. Motion during a true beat can differ from motion
during an artifact, so the two class covariances need not be equal. Let $R_0$ and $R_1$ be the
disturbance covariances under $H_0$ and $H_1$, with average $\bar{R}=\tfrac12(R_0+R_1)$. The
optimal test is then quadratic, and no single deflection fixes its operating curve. The
Bhattacharyya distance
\begin{equation}
\beta=\tfrac18\,\underline{\mu}^{*}\bar{R}^{-1}\underline{\mu}
+\tfrac12\ln\frac{|\bar{R}|}{\sqrt{|R_0|\,|R_1|}}
\end{equation}
still orders the sensor sets. With equal priors, the error probability of the optimal test is at
most $\tfrac12 e^{-\beta}$. For $R_0=R_1=R$ the second term vanishes and $\beta=d^{2}/8$. The
weighted detector below uses $\bar{R}$ in place of $R$, which keeps the first term of $\beta$. In
the recordings below, the criterion gives a predicted ranking, and the measured admission of each
sensor set tests that prediction.

Array design follows the same rule. In a linear array the difference co-array is the set of
distinct spacings between element pairs. It fixes which spatial correlations the array can form,
and a repeated spacing adds no new correlation. Minimum-redundancy arrays place few elements so
that the co-array is contiguous with few repeats~\cite{moffet1968}. With covariance augmentation,
such an array of $M$ elements resolves more than the $M-1$ uncorrelated sources of a conventional
array~\cite{pillai1985}. A constructive placement keeps the squared element count below three
times the array span~\cite{pearson1990}. Nested and coprime arrays reach a number of distinct
spacings of the order of the square of the element count~\cite{pal2010,vaidyanathan2011}. In
imaging, the difference co-array governs incoherent imaging and the sum co-array governs coherent
imaging~\cite{hoctor1990,ahmad2001}. The geometry fixes the co-array, and the weighting then
selects one point-spread function~\cite{hoctor1990}. Sensor selection and weighting split the
same way here. The chosen sensors fix $\underline{\mu}$ and $R$, and the weight
$R^{-1}\underline{\mu}$ then selects the detector. The metric there is a count of distinct
spacings. The metric here is the deflection increment. In both, a sensor belongs where it supplies
what the others do not.

\section{Data and Evaluation Protocol}

Three public resources are used. CEBS~\cite{cebs} gives ECG, respiration, and SCG from 20
volunteers at rest. EPHNOGRAM~\cite{ephnogram} gives ECG and PCG from healthy adults at rest and
during exercise. FOSTER~\cite{foster,fosterdata} gives ECG, SCG, heart sound, two
forcecardiography sensors, and respiration from 40 healthy adults at rest, 20 of them women. All
recordings are public and de-identified and were collected under the ethics approvals of the
original studies. The authors had no contact with the participants. The
same single-channel settings are used for all three, except where the multi-sensor paragraph
below says otherwise.

\textit{Records:} CEBS, from PhysioNet~\cite{physionet}, contributes its twenty basal recordings, one per volunteer, with lead~II
and the seismocardiogram (SCG). These are the files a001 to a020 of the corrected release, since the
files b002 to b020 of the current release repeat the post-music recordings. EPHNOGRAM contributes the exercise records whose ECG and
phonocardiogram (PCG) quality notes both begin with ``Good.'' Two of these duplicate other records sample for
sample and were removed, leaving 23 records of 30~min from 12 subjects (ECGPCG0029, 0031--0034,
0036, 0042, 0043, 0047, 0050, 0053, 0055, 0056, 0059--0062, 0064--0069). One retained record
carries the note ``Good; Abnormal ECG.'' Removing it changes the exercise result by less than half a
percentage point. All channels are decimated with anti-aliasing filters to 500~Hz, or 2000~Hz
for the phonocardiogram, and kept in integer units. The software and the microcontroller
therefore receive identical streams. FOSTER contributes all 40 recordings, with lead~II, the
SCG, the heart sound, and both forcecardiography sensors. Each lasts about 7~min, is taken
seated, and ends with short inspiratory and expiratory apneas. The two forcecardiography sensors are
piezoelectric, one of lead zirconate titanate (PZT) and one of polyvinylidene fluoride (PVDF).
Its five channels are decimated from 10~kHz to 500~Hz, the heart sound included.

\textit{Detectors:} The ECG detector follows the design of Pan and Tompkins~\cite{pan1985}. The
ECG is band-passed at 5--20~Hz by a causal second-order filter, differenced, squared, and averaged
over 80~ms. A local maximum of this energy is accepted as a beat when it lies one quarter of the
way above the running noise level toward the running signal level and follows the previous beat
by more than 200~ms. The signal level is seeded at half the 99th percentile of the energy in the
first 5~s, and the noise level at one tenth of that. Each level then moves one eighth of the way
toward each accepted beat or rejected candidate. The R time is the largest band-passed
value just before the energy peak. The mechanical feature is the rectified band-passed signal
averaged over 20~ms. The band is 5--40~Hz for the seismocardiogram and the two forcecardiography
sensors, and 20--150~Hz for the heart sound.

\textit{Calibration and scoring:} The electromechanical delay is learned on the first 40~s of each
record. For each beat in that segment, the lag from the R time to the largest mechanical value is
found within 250~ms, or within 150~ms when the median RR interval in the first 40~s is below
500~ms. The delay is the
upper median of these lags, and the confirmation window is the delay $\pm40$~ms. The reference
amplitude is the upper median of the mechanical peak in that window. The ECG filter lags more than
the mechanical filter, so the learned delays are about 20 to 30~ms shorter than the physiological
ones. After 40~s, an electrical event is admitted when the mechanical peak inside its window is at
least $k$ times the reference. Sweeping $k$ from 0 to 0.8 traces the operating curve.

\textit{Artifact challenge:} Transients are injected into the ECG only, and the mechanical channel
is never corrupted. Each transient is a QRS-like biphasic derivative-of-Gaussian pulse, 80~ms long, with
peak amplitude 0.75 times the record's median R amplitude. A pulse is placed in a random 20\% of
the RR gaps after 40~s, with its center at least 250~ms from either beat and a fixed random seed
per record. The transient is designed to be accepted by a single-channel ECG detector, so the
ECG alone sets the baseline against which corroboration is measured.
Beats detected on the clean ECG form the reference. Retention is the fraction of these reference
beats admitted. The admission rate is the fraction of injected transients admitted. Both use a
60~ms matching tolerance. In CEBS, retention measured against the database's own beat annotations equals retention
measured against the detector, so this reference does not favor the detector. EPHNOGRAM has no beat
annotations, and its retention is measured against the detector's clean-ECG output. Confidence intervals resample subjects, not beats, with 2000 bootstrap draws.

\textit{Time-shift control:} The control reads the mechanical feature 20~s after the electrical
event, many beats later, with the same window and reference. The second channel stays active,
but the physiological timing is broken. The last 20~s of each record are not scored, so the fused
detector and the control score the same events.

\textit{Multi-sensor fusion on FOSTER:} Each mechanical sensor has its own learned delay, window,
and reference. Two rules combine the sensors. The AND rule admits an event when every sensor in
the set confirms it. The vote admits an event when all but one sensor confirm it, and applies to sets of three or more sensors. For each of the
15 sets that can be formed from the four sensors, $k$ is swept from 0 to 1.5, and admission is read
at 95\% retention. The threshold is also chosen on held-out subjects. For each subject, $k$ is the
largest value that keeps 95\% retention on the other 39. Two extra challenges test when more
sensors help. In the first, each sensor independently receives a mechanical burst at its learned
delay after a detected transient, with probability 0.5. In the second, every sensor receives the
burst. The burst is a band-limited pulse whose feature peak equals the sensor's reference.
Differences between sets are paired. Every set uses the same 2000 subject resamples.
The weighted detector of Section~\ref{sec:criterion} takes as the confirmation feature of each
sensor the logarithm of the ratio between its window peak and its reference. It admits an event
when $\underline{\mu}^{*}R^{-1}\underline{x}$ exceeds a threshold. For each subject,
$\underline{\mu}$ and $R$ are estimated from the other 39 subjects. The class covariances of beats
and transients need not be equal, so $R$ is their average $\bar{R}$ of Section~\ref{sec:criterion}. The weighted scores of all held-out
subjects are pooled, and admission is read at 95\% pooled retention. The weights are therefore
held out by subject. The operating threshold is read from the pooled curve. A split-half check
fixes both the weights and the threshold on 20 subjects and scores the other 20, over 200 random
splits. The deflection of each set is estimated on all 40 subjects.

\textit{Fixed-point replay:} The same detector is written in C with 64-bit integer arithmetic, and
the confirmation test is an exact integer comparison. The code learns the delay and reference on
the first 40~s of the stream. One source file builds both the ESP32 firmware and a host program.
The replay streams the same integer records through the chip, 0.5~s at a time, and compares every
decision with the floating-point software. It tests whether the device reproduces the software and
keeps pace with the stream.

\section{Results}
\begin{table}[!t]
\caption{Admission of injected transients at 95\% clean-beat retention, in percent. The timed
sensor is the SCG at rest and the PCG during exercise. The ECG alone has no mechanical
threshold. Its row is the fraction of transients its detector accepted.}
\label{tab:summary}
\centering\small
\begin{tabular}{@{}lccc@{}}
\hline
Detector & CEBS & EPHNOGRAM & FOSTER \\
\hline
ECG alone & 100 & 99.8 & 96.8 \\
One timed sensor & 5.6 & 34.8 & 12.3 \\
AND of four & -- & -- & 10.4 \\
Three-of-four vote & -- & -- & 8.1 \\
Weighted pair & -- & -- & 6.2 \\
Weighted, all four & -- & -- & 6.7 \\
\hline
\end{tabular}
\end{table}
Table~\ref{tab:summary} collects the main result. One timed mechanical sensor removes most
injected transients, and complementary sensors remove more.
\begin{figure}[!t]\centering
\begin{subfigure}{0.47\columnwidth}\centering
\includegraphics[width=\linewidth]{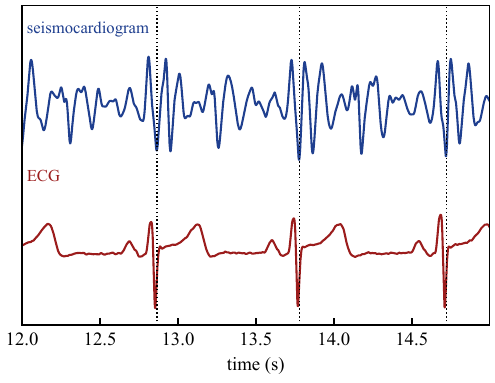}\caption{}
\end{subfigure}\hfil
\begin{subfigure}{0.47\columnwidth}\centering
\includegraphics[width=\linewidth]{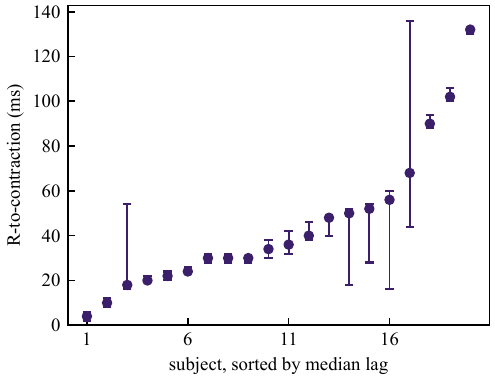}\caption{}
\end{subfigure}
\caption{Electromechanical timing in CEBS. (a)~Seismocardiogram (top) and ECG (bottom) of
record a001 from 12~s. Dotted lines mark the R peaks. (b)~Median R-to-contraction lag per subject
with interquartile bars, sorted by median. Lags are read after the detection filters and run
about 20 to 30~ms short of the physiological delay.}
\label{fig:emcoup}
\end{figure}
The physiological timing needed by the corroboration rule is visible directly in the CEBS
recordings (Fig.~\ref{fig:emcoup}). The lag is tight within most subjects, but the median lag
ranges from a few milliseconds to more than 100~ms across subjects. A fixed population window
would either miss subjects with a longer delay or admit broad timing in subjects with a shorter
one, so the detector learns the interval per subject. In 19 of the 20 subjects the delay learned
in the first 40~s lies within 4~ms of the median lag over the whole record. In the remaining
subject the calibration locked onto a different peak of the seismocardiogram.

\begin{figure}[!t]\centering
\begin{subfigure}{0.47\columnwidth}\centering
\includegraphics[width=\linewidth]{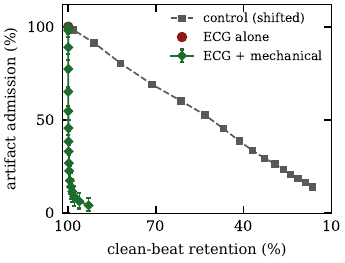}\caption{}
\end{subfigure}\hfil
\begin{subfigure}{0.47\columnwidth}\centering
\includegraphics[width=\linewidth]{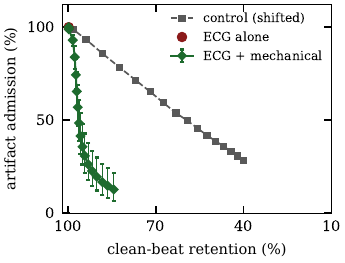}\caption{}
\end{subfigure}
\caption{Synthetic-artifact trade-off. (a)~CEBS at rest, ECG and seismocardiogram, 20 subjects.
(b)~EPHNOGRAM during exercise, ECG and phonocardiogram, 12 subjects. Bars are 95\% intervals
from resampling subjects. The shifted curve reads the mechanical signal 20~s late.}
\label{fig:emtrade}
\end{figure}

Figure~\ref{fig:emtrade} is the main test of the fusion mechanism, and
Table~\ref{tab:ops} lists its operating points. The ECG alone admits essentially every injected
transient. At rest the timed seismocardiogram removes nineteen in twenty of them at 95\%
retention. During exercise the phonocardiogram removes about two thirds of them at the same
retention, so the rejection is partial. When the mechanical data are shifted by 20~s, transients
are admitted about as often as beats, and the gate no longer separates them. The benefit is not
reproduced by the shifted control, so it requires electromechanical timing and not merely an
active second channel. The timed gate admitted fewer transients than the shifted control in every
subject of both datasets. The thresholds in Table~\ref{tab:ops} were read from the curves. When
the threshold is instead chosen on half of the subjects and applied to the other half, over 300
random splits and with the control's threshold chosen the same way, the timed gate admits 6.8\% of the transients at 95.8\% retention at rest, and 51.5\% at 93.4\% retention
during exercise, while the shifted control admits more than 95\% in both. The threshold therefore
carries over between subjects at rest but only in part during exercise. A fixed window at the
population delay also admitted slightly fewer transients during exercise than the per-subject
window, 30.6\% against 34.8\% at 95\% retention, whereas at rest the per-subject window was
better, 5.6\% against 7.5\%.

\begin{table}[!t]
\caption{Operating points of Fig.~\ref{fig:emtrade}. Intervals for the fused detector are
95\% intervals from resampling subjects. CEBS scores 5619 beats and 1124 transients; EPHNOGRAM
scores 76\,416 beats and 8355 transients.}
\label{tab:ops}
\centering\small
\begin{tabular}{@{}lccc@{}}
\hline
Detector & $k$ & Retention (\%) & Admitted (\%) \\
\hline
\multicolumn{4}{@{}l}{\textit{CEBS, seismocardiogram, rest}} \\
ECG alone & -- & 100 & 100 \\
Fused & 0.6 & 99.1 & 14.2 (7.8--21.6) \\
Fused & 0.8 & 93.1 & 4.3 (1.3--8.2) \\
Shifted control & 0.6 & 26.4 & 23.4 \\
\hline
\multicolumn{4}{@{}l}{\textit{EPHNOGRAM, phonocardiogram, exercise}} \\
ECG alone & -- & 99.9 & 99.8 \\
Fused & 0.5 & 94.3 & 30.8 (21.6--42.8) \\
Fused & 0.8 & 84.5 & 12.8 (6.6--21.6) \\
Shifted control & 0.5 & 55.8 & 45.6 \\
\hline
\end{tabular}
\end{table}

\begin{figure}[!t]\centering
\begin{subfigure}{0.47\columnwidth}\centering
\includegraphics[width=\linewidth]{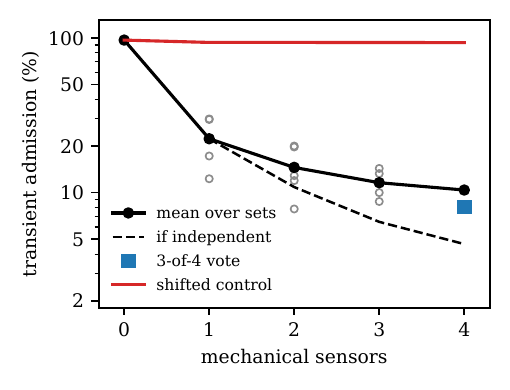}\caption{}
\end{subfigure}\hfil
\begin{subfigure}{0.47\columnwidth}\centering
\includegraphics[width=\linewidth]{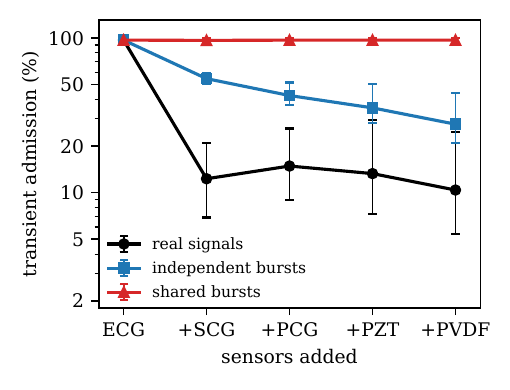}\caption{}
\end{subfigure}
\caption{More mechanical sensors on FOSTER, 40 subjects at rest, admission at 95\% clean-beat
retention. (a)~AND rule. Gray circles are the sensor sets and the black line their mean. The
dashed line is the admission if the sensors failed independently. The square is the
three-of-four vote. (b)~Sensors added in the order SCG, heart sound, PZT, PVDF, with the real
signals, a burst in each sensor with probability 0.5, and a burst in every sensor. Bars are
95\% intervals from resampling subjects.}
\label{fig:foster}
\end{figure}

FOSTER tests whether more sensors lower admission further (Fig.~\ref{fig:foster}). The ECG alone
admitted 96.8\% of the transients. With one mechanical sensor, admission averaged 22.3\% over the
four sensors, and the seismocardiogram alone gave 12.3\%. Averaged over sensor sets, admission fell
to 14.5\% with two sensors, 11.6\% with three, and 10.4\% with four. Each added sensor gained less
than the one before.
The sensors did not fail independently. With four sensors, independent failures would give 4.7\%,
less than half the measured value. A transient is likely admitted when its window falls on real
cardiac motion, which every sensor records. As a result, the AND rule with all four sensors did not
differ significantly from the seismocardiogram alone. The three-of-four vote did modestly better. It
admitted 8.1\% of the transients at 95\% retention. With the threshold chosen on held-out subjects, it admitted 4.4 points fewer than the
seismocardiogram alone, with a paired 95\% interval of $-9.5$ to $-0.2$ points. It admitted fewer
transients in 27 of the 40 subjects. The vote tolerates one sensor that misses a real beat, so its
threshold can stay stricter at the same retention.

The two extra challenges show when more sensors help (Fig.~\ref{fig:foster}(b)). With the real
signals, adding the heart sound to the seismocardiogram raised admission slightly. The AND rule
must loosen every threshold to keep 95\% joint retention, and the heart sound adds little
independent check. With independent
bursts, admission fell at every step, from 54.5\% with one sensor to 27.7\% with four. With a burst
in every sensor, admission stayed near 97\% however many sensors were used. The shifted control
stayed near 93\% at every number of sensors. More sensors therefore lower admission only when they
add independent, correctly timed checks. Leaving out four subjects with ECG detection problems gave
the same pattern.

\begin{figure}[!t]\centering
\begin{subfigure}{0.47\columnwidth}\centering
\includegraphics[width=\linewidth]{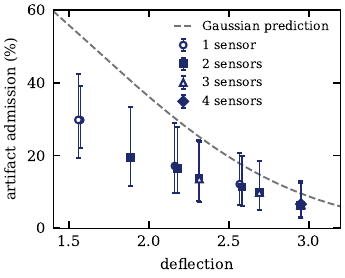}\caption{}
\end{subfigure}\hfil
\begin{subfigure}{0.47\columnwidth}\centering
\includegraphics[width=\linewidth]{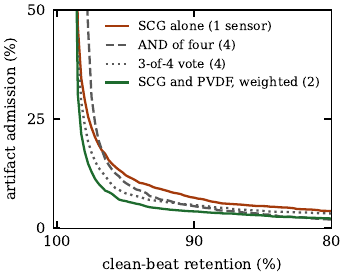}\caption{}
\end{subfigure}
\caption{Weighted fusion on FOSTER, weights held out by subject. (a)~Admission at 95\%
retention against the deflection of the 15 sensor sets. Markers give the sensor count, bars are
95\% intervals from resampling subjects, and the dashed line is the Gaussian prediction.
(b)~Operating curves of four detectors. The number in parentheses is the sensor count.}
\label{fig:fosterw}
\end{figure}

Fig.~\ref{fig:fosterw} applies the criterion of Section~\ref{sec:criterion} to the same FOSTER
events. In Fig.~\ref{fig:fosterw}(a), a larger deflection goes with a lower held-out admission
across all 15 sensor sets, with a Spearman rank correlation of $-0.96$. The agreement in rank is
close but not exact. The deflections use all 40 subjects, so the ranking is within the cohort
and is not a prospective test. The Bhattacharyya distance of Section~\ref{sec:criterion}, which
lets the two class covariances differ, ranks the sets with a correlation of $-0.90$ against
admission. The Gaussian prediction follows the ranking but overstates the admission of
every set, so the absolute rate must still be read from recordings. Given the seismocardiogram,
the PVDF forcecardiography sensor adds 2.10 to the squared deflection, the PZT sensor adds 0.64
and the heart sound adds 0.08. The heart sound is a useful sensor alone, but the seismocardiogram
already predicts most of its mean shift.
Fig.~\ref{fig:fosterw}(b) shows the consequence. The weighted pair of the seismocardiogram and the
PVDF sensor admitted 6.2\% of the transients at 95\% retention, about half the admission of the
seismocardiogram alone, with a paired 95\% interval of $-11.0$ to $-2.3$ points. With the pair, the
weights, and the threshold all chosen on half of the subjects, the pair admitted 5.5\% of the
transients on the other half at 93.7\% retention, against 10.5\% for the seismocardiogram alone.
The same pair was chosen in every one of 400 training halves. Adding the other two sensors to the pair changed its admission by
less than half a point. The weighted pair also admitted fewer
transients than the AND rule of all four sensors and did not differ significantly from the
three-of-four vote.

\begin{figure}[!t]\centering
\includegraphics[width=0.6\columnwidth]{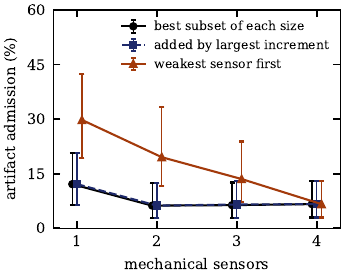}
\caption{Admission of injected ECG transients at 95\% clean-beat retention against the number of
mechanical sensors for the weighted detector on FOSTER, under three orders of adding sensors.
Bars are 95\% intervals from resampling subjects.}
\label{fig:orders}
\end{figure}

Fig.~\ref{fig:orders} shows the weighted detector against the number of sensors under three
orders. The best subset of each size falls from 12.1\% with one sensor to 6.2\% with two and
stays within half a point after that. Adding sensors by largest increment follows the same path. Starting from the weakest sensor, the PZT at
29.8\%, each added sensor lowers admission, to 19.5\%, 13.5\% and 6.7\%. All three orders end at
the same four-sensor point.

\begin{figure*}[htb]\centering
\begin{subfigure}[b]{0.225\textwidth}\centering
\includegraphics[width=\linewidth]{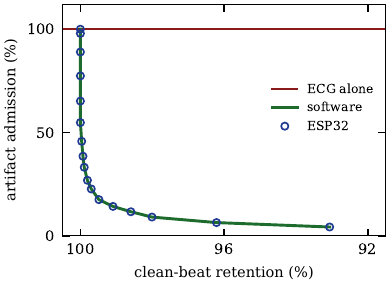}\caption{}
\end{subfigure}\hfil
\begin{subfigure}[b]{0.225\textwidth}\centering
\includegraphics[width=\linewidth]{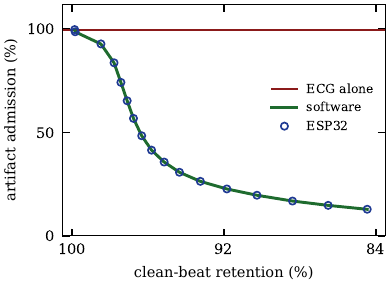}\caption{}
\end{subfigure}\hfil
\begin{subfigure}[b]{0.175\textwidth}\centering
\includegraphics[width=\linewidth]{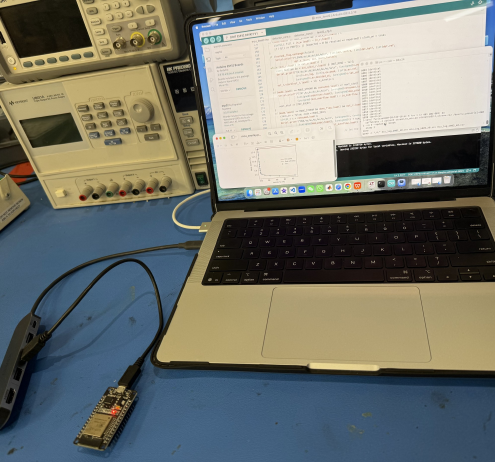}\caption{}
\end{subfigure}
\caption{Fixed-point replay of the experiment in Fig.~\ref{fig:emtrade}, same records,
injections, and windows. (a)~CEBS: the ESP32 decisions (circles) follow the software curve.
(b)~EPHNOGRAM: the same agreement during exercise. (c)~The replay bench. The embedded experiment
verifies the fusion detector only.}
\label{fig:mcu}
\end{figure*}

The ESP32 replay reproduces the software (Fig.~\ref{fig:mcu}). In CEBS no decision differs at any
threshold. In EPHNOGRAM 20 of about 1.45 million decisions differ, at most four at any threshold,
all near the threshold and caused by fixed-point rounding. The chip's decisions match those of the
same C code compiled on a PC, as expected for integer arithmetic. Table~\ref{tab:esp32} gives the
build and timing. The detector needs about 4~ms of computation per beat, far less than the shortest
beat interval. Once per record, the end of the 5~s seed segment takes 26~ms in a single tick. The
stream arrives in 0.5~s blocks, so this delay is absorbed by the input buffer.

\begin{table}[!t]
\caption{Fixed-point detector on the ESP32 (Arduino-ESP32 core 3.3.5, target \texttt{esp32},
optimization \texttt{-Os}).}
\label{tab:esp32}
\centering\footnotesize
\begin{tabular}{@{}lp{0.62\columnwidth}@{}}
\hline
Chip and clock & ESP32-D0WD-V3 (rev.~3.1, Xtensa LX6, dual core), 240~MHz, 4~MB flash \\
Arithmetic & 64-bit integer fixed point; no floating point in the detector \\
Flash (sketch) & 283\,383~B of 1\,310\,720~B (21\%) \\
RAM & 83\,968~B static (25\% of 327\,680~B); the learning buffers are on the heap, with 151\,464~B of heap still free at the end of a record \\
Time per tick & mean 11.2~$\mu$s (CEBS, one mechanical sample per tick) and 15.1~$\mu$s (EPHNOGRAM, four); largest tick outside the seed segment 0.8~ms (CEBS) and 0.9~ms (EPHNOGRAM); once per record a 26~ms tick sorts the 5~s seed segment of the ECG threshold \\
Time per beat & all ticks between consecutive beats: median 4.3~ms and 99th percentile 6.0~ms in CEBS (7282 beats, three identical runs); median 3.5~ms and 99th percentile 6.0~ms in EPHNOGRAM (85\,505 beats) \\
\hline
\end{tabular}
\end{table}

\section{Discussion and Limitations}
The strongest mechanistic result is the time-shift control. It tests a direct alternative
explanation for the fusion curve, namely that any additional channel would reject the injected
ECG transients. The shifted control preserves a busy mechanical signal but breaks the expected
physiological delay. The shifted control does not reproduce the benefit. That is evidence for a
timing-specific corroboration mechanism rather than a generic two-channel effect. FOSTER adds a
second finding. An added sensor lowers admission by the part of its response that the sensors in
place cannot predict, and the deflection increment measures that part before any operating curve
is drawn. The heart sound, useful alone, earns almost nothing next to the seismocardiogram, while
a forcecardiography sensor halves its admission. Every order of adding sensors ends at the same
floor, and the increment reaches it with two sensors. The FOSTER recordings are at rest, so this
ranking of sensors remains to be tested during exercise.

The main limitation is the kind of artifact tested. The injected transients are placed in the
ECG alone, whereas natural ambulatory motion can contaminate both the electrical and mechanical
channels. The
retention $P_D$ counts retained clean-ECG candidates, not detected clinical events. The admission
$P_F$ counts admitted injected transients, not false alerts per hour of wear. The gate only removes
candidates. It cannot recover a beat that the ECG detector never proposed, so no higher absolute
beat sensitivity is shown here. Testing that claim needs ECG-only and fused detectors compared
at matched false-alarm burden against independent annotations on held-out subjects. If realistic motion produces correlated
false events in both modalities at plausible delays, the present advantage could shrink or
disappear. The shared-burst challenge on FOSTER
shows this worst case. When every sensor receives the burst at its learned delay, no number of
sensors helps. A decisive next experiment therefore
requires naturally occurring motion with beat annotations.
The exercise result is also weaker than the resting one. The phonocardiogram gate removes about two
thirds of the transients at 95\% retention, against nineteen in twenty at rest, and a threshold
chosen on other subjects carries over only in part. A plausible reason is that the 80~ms window
around a delay learned in the first 40~s does not follow the shortening of the electromechanical
interval as heart rate rises. A fixed population window did no worse than the learned one during
exercise, which agrees with this reading but does not test it. The challenge also favors
the gate. Each transient center lies at least 250~ms from a beat, and the mechanical channel stays
clean. During exercise this guard leaves no room for a transient when the RR interval is below
500~ms, so the stretches with heart rate above 120 beats per minute carry no transients.
In two exercise records the ECG detector found no beat for about 400~s after saturation
artifacts, although an ECG was present. These stretches carry no scored events.
EPHNOGRAM records were selected by good quality notes and come from healthy men aged 23 to 29.
FOSTER adds 20 women, but only at rest. Unselected noisy records and exercise recordings in women remain untested.

All three resources contain healthy volunteers, so the detector has not been tested for adverse-event
prediction. The proposed architecture also allows a second, contextual check on the phone, for
example a compact physiological model that interprets a confirmed event. That stage is not
built or tested here.
Only the chest electrical and mechanical pair is validated. The other sites in
Fig.~\ref{fig:overview} are untested extensions. A foot pulse timed against the chest ECG gives a
pulse-arrival time, which differs from pulse-transit time by the pre-ejection
period~\cite{zhang2011,mukkamala2015}. An ear pulse sensor, possibly in the temple tips of
eyeglasses~\cite{holz2017}, could give a shorter timing path. Auricular vagal stimulation has been
studied in humans~\cite{yakunina2017,clancy2014}, but its placement, dose, and safety are outside
this work. A phone pressed to the chest could also serve as an occasional sensor, with its
microphone and accelerometer as the mechanical channels.

\section{Conclusion}
In the clinic a physician corroborates the electrocardiogram. A wearable has no physician, and
the results support physiologically timed mechanical confirmation as its substitute, a
first-stage check for a phone-centered monitor. They do not validate clinical alert decisions. In the Gaussian model, an
added channel helps only through the part of its response that the channels already in place
cannot predict. On the two single-sensor datasets, an electrical event confirmed by a mechanical
contraction at the subject's own delay rejected most injected ECG-only transients while keeping
nearly all clean beats. The rejection was strong at rest and partial during exercise. With four
mechanical sensors at rest, the deflection of the joint confirmation features closely tracked the
ranking of the sensor sets by measured admission, and the pair it selected halved the admission
of the best single sensor. A time shift of the mechanical channel removed the
separation, so within this synthetic-artifact challenge the benefit rests on timing. The detector
runs in fixed-point arithmetic on an ESP32 at a few milliseconds per beat and reproduces all but a
few near-threshold software decisions.
The next tests are specific. They are natural motion that disturbs both channels, thresholds
fixed on separate subjects with end-to-end detection reported at matched artifact admission, a
confirmation window
that follows the heart rate during exercise, cohorts with labeled clinical events, and the
runtime and power of the complete pipeline on a phone.

\section*{Acknowledgment}
AI tools were used for editorial assistance, proofreading, and language refinement. All
theoretical developments, derivations and figures are the authors' work. The CEBS~\cite{cebs}, EPHNOGRAM~\cite{ephnogram},
and FOSTER~\cite{foster,fosterdata} recordings behind Table~\ref{tab:summary} are public
data. FOSTER carries a Creative Commons Attribution license.

\begin{IEEEbiographynophoto}{Zixuan Fu}
received her B.S. degree in electrical engineering from Tianjin University, Tianjin,
China, in 2024 and her M.S. degree in electrical engineering from New York University,
New York, NY 11201 USA, in 2026, where she is currently a research intern. Her research
interests include digital communications, machine learning, signal processing for
biomedical applications. She is a Student Member of
IEEE.
\end{IEEEbiographynophoto}

\begin{IEEEbiographynophoto}{Anthony Raffaele}
is currently working toward his B.S. degree in electrical engineering (expected 2027) at New York
University, New York, NY 11201 USA. His research
interests include embedded and low-power hardware, wearable sensing systems, and
real-time signal processing for biomedical applications. He is a Student Member of IEEE.
\end{IEEEbiographynophoto}

\begin{IEEEbiographynophoto}{Unnikrishna Pillai}
received his Ph.D. degree in systems engineering from the University of Pennsylvania in
1985. He joined Brooklyn Polytechnic, New York, in 1985 as an assistant professor in
electrical engineering and since 1995 has been a professor of electrical and computer
engineering, now the Department of Electrical and Computer Engineering, New York
University, Brooklyn, NY 11201 USA. He has authored or coauthored several widely used
textbooks, among them \emph{Probability, Random Variables and Stochastic Processes}
(A. Papoulis and S. U. Pillai, 2002). His research interests include synthetic aperture radar imaging, rational approximation, array signal
processing and applying nonlinear equations to physical
systems. He is a Fellow of IEEE.
\end{IEEEbiographynophoto}

\end{document}